\documentclass[conference]{IEEEtran}
\IEEEoverridecommandlockouts
\usepackage{cite}
\usepackage{balance}
\usepackage{easybmat}
\usepackage{stfloats}
\usepackage{amsmath,amssymb,amsfonts}
\usepackage{graphicx}
\usepackage{textcomp}
\usepackage[linesnumbered,lined,boxed,commentsnumbered, ruled]{algorithm2e}
\usepackage{epstopdf}
\usepackage[printonlyused,withpage]{acronym}
\usepackage{caption}
\usepackage{subcaption}
\usepackage{algorithmic}
\usepackage{multirow}

\usepackage[letterpaper, left=0.7in, right=0.7in, bottom=1in, top=0.7in]{geometry}

\newtheorem{prop}{Proposition}

\newtheorem{obs}{Observation}

\newenvironment{proof}{ \paragraph*{\hspace{-1em}Proof}}{\hfill$\square$}
\renewcommand{\baselinestretch}{.97} 

\usepackage{xcolor}

\newcommand{\bs}[1]{\ensuremath{\boldsymbol{#1}}}
\newcommand{\mrm}[1]{\ensuremath{\mathrm{#1}}}
\renewcommand{\vec}[1]{\ensuremath{\mathrm{vec}}}

\newcommand{\diag}{\mathrm{diag}} 

\newcommand{\comment}[1]{}

\acrodef{MU-MIMO}{multi-user multiple-input multiple-output}
\acrodef{UE}{user equipment}
\acrodef{BS}{base station}
\acrodef{ZF}{zero-focing}
\acrodef{NLA-ZF}{non-linearity-aware \ac{ZF}}
\acrodef{CSI}{channel state information}
\acrodef{Tx-chain}{transmit-chain}
\acrodef{OTA}{over-the-air}
\acrodef{TDD}{time division duplex}
\acrodef{SNR}{signal-to-noise ratio}
\acrodef{SIR}{signal-to-interference ratio}
\acrodef{SDR}{signal-to-distortion ratio}
\acrodef{SNDR}{signal-to-noise-plus-distortion ratio}
\acrodef{SINDR}{signal-to-interference-noise-and-distortion ratio}
\acrodef{RHS}{right-hand side}
\acrodef{LHS}{left-hand side}
\acrodef{PA}{power amplifier}
\acrodef{DPD}{digital pre-distortion}

\begin{document}

\title{Zero-Forcing MU-MIMO Precoding under\\ Power Amplifier Non-Linearities}
\author{Juan Vidal Alegr\'{i}a, Ashkan Sheikhi, and Ove Edfors\\
\IEEEauthorblockA{Department of Electrical and Information Technology, Lund University, Lund, Sweden\\} 
\{name.surname(\_surname2)\}@eit.lth.se}

\maketitle

\begin{abstract} 
In  multi-user multiple-input multiple-output (MU-MIMO) systems, the non-linear behavior of the power amplifiers (PAs) may cause degradation of the linear precoding schemes dealing with interference between user equipments (UEs), e.g., the zero-forcing (ZF) precoder. One way to minimize this effect is to use digital-pre-distortion (DPD) modules to linearize the PAs. However, using perfect DPD modules is costly and it may incur significant power consumption. As an alternative, we consider the problem of characterizing non-linearity-aware ZF (NLA-ZF) precoding schemes, hereby defined as linear precoders that achieve perfect interference cancellation in the presence of PA non-linearity by exploiting knowledge of this non-linear response. We provide initial iterative solutions that allow achieving NLA-ZF (up to adjustable tolerance) in a two-UE downlink MU-MIMO scenario where the base station (BS) has an even number of antennas, and each antenna is connected to a PA exhibiting third-order memory-less non-linear behavior. The proposed approach  allows for performance gains in scenarios with significant residual interference.
\end{abstract}

\begin{IEEEkeywords} Power amplifier (PA) distortion, MU-MIMO precoding, Non-linearity-aware zero-forcing (NLA-ZF).
\end{IEEEkeywords}
\section{Introduction}
Amplifiers deployed in the \ac{Tx-chain} of wireless communication systems are commonly operating in the non-linear regime to maximize energy-efficiency \cite{schenk_rf}. On the other hand, \ac{MU-MIMO} systems \cite{mu-mimo} have become a pivotal technology in current wireless systems, specially considering its scaled-up version, massive MIMO \cite{marzetta}, which has been a key enabling technology for the development of 5G \cite{mMIMO5G}. The core benefits of these systems come from their ability to improve spectral efficiency by multiplexing \acp{UE} in the spatial domain \cite{mimo}.

In order to spatially multiplex multiple single-antenna \acp{UE}, the \ac{BS} should precode the symbols intended to each \ac{UE} to compensate for the MIMO channel. Linear precoders are most desirable due to their relatively low complexity, while allowing close-to-optimal performance when scaling up the number of \ac{BS} antennas \cite{rusek}. One common linear precoding strategy is \ac{ZF}, which can effectively remove the multi-user interference by inverting the MIMO channel. However, due to the non-linearity of the \acp{PA}, the \ac{ZF} precoder may not be able to perfectly remove the interference of the MIMO channel, as will be shown in this work.

A conventional way of mitigating amplifier distortion is to employ a \ac{DPD} module that allows transforming the transmitted signals in the digital domain by pre-inverting the amplifier non-linear response that they will experience \cite{dpd_morgan}. A perfect \ac{DPD} module theoretically allows for perfect linearization of an arbitrary non-linear response, which would allow to perfectly exploit the benefits of traditional linear precoding approaches such as \ac{ZF}. However, these \ac{DPD} modules require high-computational complexity, incurring significant power consumption \cite{DPDcomplexity1,DPDPower_Wes,DPDPower_Wu}. Note that, even for a third-order non-linearity, the number of \ac{DPD} coefficients required to provide a perfect inversion is theoretically unbounded. On the other hand, using low-end \ac{DPD} modules, which allows reducing cost and energy consumption, may still incur residual non-linear behavior \cite{muris_pw_sc}.

In this work, we study linear precoding approaches with the goal of achieving perfect interference cancellation in the presence of \ac{PA} non-linearity. We use the term \ac{NLA-ZF} to refer to such precoding strategies since they employ the knowledge of the \ac{PA} non-linear response to attain the \ac{ZF} goal. Note that, the knowledge of the \ac{PA} amplifier response for each \ac{Tx-chain} may be estimated in practice using \ac{OTA} methods, e.g., relying on inter-antenna coupling measurements as in \cite{ota_dpd}. As an initial proof of concept, we derive concrete \ac{NLA-ZF} approaches for third-order \ac{PA} non-linearity in a two-\ac{UE} scenario with an even number of \ac{BS} antennas, while future work may consider how to extend these results to more general scenarios. The considered approach provides a cost- and energy-efficient solution to deal with multi-user interference which does not rely on perfect \ac{DPD} modules.

\section{System Model and Problem Formulation}
We consider a narrowband \ac{TDD} \ac{MU-MIMO} scenario where an $M$ antenna \ac{BS} serves $K$ single-antenna \acp{UE} in the downlink. The vector of complex baseband symbols received by the \acp{UE} may be expressed as
\begin{equation}\label{eq:mimo_nl}
    \bs{y} = \bs{H}\bs{f}(\bs{W}\bs{s})+\bs{n},
\end{equation}
where $\bs{H}$ is the $K \times M$ channel matrix, which is assumed to be full-rank and perfectly known via uplink pilots, $\bs{n}\sim CN(\bs{0},N_0\mathbf{I}_K)$ is the additive white complex-Gaussian noise vector, $\bs{s}\sim CN(\bs{0},\mathbf{I}_K)$ is the $K\times 1$ vector of complex-Gaussian symbols containing the information for each user, $\mathbf{I}_K$ denotes the $K \times K$ identity matrix, $\bs{W}$ is the $M\times K$ linear precoder, and $\bs{f}(\cdot)$ is a vector valued function capturing the baseband-equivalent non-linear response of the \acp{PA}. Since the third order terms are the main source of in-band amplifier distortion \cite{nl_thrd_ord}, and assuming that coupling between \acp{Tx-chain} associated to different antennas is negligible, we model $\bs{f}(\cdot)$ as a component-wise function where each entry is given by a third-order memoryless polynomial \cite{schenk_rf}
\begin{equation}\label{eq:nl_thr_ord}
    f_m(x)=a_{1,m} x+a_{3,m} \vert x\vert^2x.
\end{equation}
We further assume $a_{1,m}$ and $a_{3,m}$, $\forall m$, known at the \ac{BS} since they may be estimated from \ac{OTA} measurements, as recently proposed in \cite{ota_dpd_nutti,ota_dpd}.

Based on the general \ac{MU-MIMO} downlink framework \cite{mimo}, we consider that the \ac{BS} may allocate power to different \acp{UE} according to pre-established system requirements. This leads to a power constraint on the columns of $\bs{W}$
\begin{equation}\label{eq:pw_cons}
    \Vert\bs{w}_k^\mrm{col}\Vert^2 = E_{s,k}, \;\;\; \forall k \in\{1,\dots, K\},
\end{equation}
where $\bs{w}_k^\mrm{col}$ corresponds to the $k$th column of $\bs{W}$, and $\sum_{k}E_{s,k}=E_s$ corresponds to the total symbol energy available at the \ac{BS}.

\subsection{Bussgang decomposition}
Employing the Bussgang theorem for the non-linearity defined in \eqref{eq:nl_thr_ord}, we can rewrite \eqref{eq:mimo_nl} as
\begin{equation}\label{eq:MIMO_Buss}
\bs{y} = \bs{H}\bs{G}(\bs{W})\bs{W}\bs{s}+\bs{H}\bs{\eta}+\bs{n},
\end{equation}
where $\bs{\eta}$ is the distortion term uncorrelated to $\bs{x}=\bs{W}\bs{s}$, and $\bs{G}(\bs{W})$ is the Bussgang gain matrix given by \cite{bussgang_emil,lis_amp_dist}
\begin{equation}\label{eq:BussG}
\begin{aligned}
\bs{G}(\bs{W})=&\diag(\alpha_1,\dots,\alpha_M)\\
&+\diag\Big(\beta_1\Vert \bs{w}_1^\mrm{T}\Vert^2,\dots,\beta_M\Vert \bs{w}_M^\mrm{T}\Vert^2\Big),
\end{aligned}
\end{equation}
with $\bs{w}_m^\mrm{T}$ corresponding to the $m$th row of $\bs{W}$, $\alpha_m\triangleq a_{1,m}$, and $\beta_m\triangleq 2a_{3,m}$. For the 3rd order non-linearity model considered in \eqref{eq:nl_thr_ord}, and based on the results from \cite{lis_amp_dist,emil_hw_dist}, the covariance matrix of $\bs{\eta}$ can be calculated as
\begin{equation}\label{Cetaeta_corr}
\begin{aligned}
	\bs{C_{\eta\eta}} =\frac{1}{2}\bs{B}(\bs{W}\bs{W}^H)\odot (\bs{W}^*\bs{W}^T) \odot (\bs{W}\bs{W}^H)\bs{B}^H,
\end{aligned}
\end{equation}
where $\odot$ denotes Hadamard product, and $\bs{B}=\diag(\beta_1,\dots,\beta_M)$. Note that this covariance matrix is generally non-diagonal. 

From the Bussgang theorem \cite{bussgang_th}, we have that that $\bs{\eta}$ and $\bs{s}$ are uncorrelated in \eqref{eq:MIMO_Buss}. Hence, the interference among \acp{UE} is essentially associated to the off-diagonal elements of the matrix product $\bs{H}\bs{G}(\bs{W})\bs{W}$.\footnote{Note that, although $\bs{s}$ and $\bs{\eta}$ may not be strictly independent, the interference is commonly measured via second-order moments, so that we may disregard higher-order dependencies.} We may now note that, when  $\bs{G}(\bs{W})$ does not correspond to a scaled identity matrix, selecting $\bs{W}$ as the traditional right pseudo-inverse of $\bs{H}$ would not achieve perfect interference cancellation. In the considered framework, this may happen for two reasons:
\begin{enumerate}
    \item Using the traditional pseudo-inverse definitions may lead to different power per antenna, i.e., we may not be able to fulfill $\Vert\bs{w}_m\Vert^2= \kappa,\; \forall m$.
    \item Even if we use the same \acp{PA} throughout all the antennas, the tolerance of the hardware components associated to them may lead to variations in the response of each \ac{Tx-chain}, i.e., we cannot assume that $\alpha_m=\alpha, \;\forall m$ and $\beta_m=\beta, \;\forall m$.
\end{enumerate}
In \cite{gen_pinv}, several generalized pseudo-inverses are defined to allow for \ac{ZF} precoding under specific constraints. The pseudo-inverse definition with per-antenna power constraints could be employed to mitigate the effect of the first issue. However, this approach only enforces inequality constraints, which is not enough to fully address this first issue since it may not generally attain $\Vert\bs{w}_m\Vert^2= \kappa,\; \forall m$ (unless power is sufficiently restricted, leading to energy-efficiency reductions as for amplifier back-off). Furthermore, the second issue would still limit the interference cancellation performance when using highly non-ideal hardware.

\subsection{Problem formulation}
The goal of this work is to find the linear precoder such that the received symbols at each \ac{UE} are not interfered by the symbols intended to other \acp{UE}. In other words, we want to find the matrices $\bs{W}$ fulfilling \eqref{eq:pw_cons} that solve
\begin{equation}\label{eq:cond_int_canc}
    \bs{H}\bs{G}(\bs{W})\bs{W} =\diag(\gamma_1,\dots,\gamma_K),
\end{equation}
where $\gamma_k$ are some non-zero arbitrary scalars. Since the distortion term $\boldsymbol{\eta}$ in \eqref{eq:MIMO_Buss} is uncorrelated to $\bs{s}$, the condition \eqref{eq:cond_int_canc} would ensure infinite \ac{SIR} at the receiver. However, as happens with traditional \ac{ZF} approaches, it may lead to noise and/or distortion enhancement effects. Note that the second order statistics of $\bs{\eta}$ do depend on $\bs{W}$. Thus, this approach is mainly suitable for interference-limited scenarios, e.g., at high \acp{SNDR}.

\section{Non-Linearity-Aware Zero-Forcing}
The main challenge towards solving \eqref{eq:cond_int_canc} comes from the dependency between $\bs{G}$ and $\bs{W}$, which is further a non-linear relation as seen in \eqref{eq:BussG}. Since we focus on interference cancellation, we may disregard the specific values of $\gamma_k$ in \eqref{eq:cond_int_canc}, associated to the resulting channel gain of user $k$. Note that the values of $\gamma_k$ would naturally become non-zero from the rank and power assumptions. We may thus focus on the non-diagonal entries in \eqref{eq:cond_int_canc}. In other words, we would like to cancel the interference of each user $i$ on all other users $k\neq i$ by solving
\begin{equation}\label{eq:int_ki}
    \sum_{m=1}^{M}h_{km}w_{mi}(\alpha_m+\beta_m\Vert\bs{w}_m^\mrm{T}\Vert^2)=0, \;\; \forall (k,i), k\neq i.
\end{equation}
A solution to \eqref{eq:int_ki}, either numerical or explicit, seems initially out of reach due to the complex interdependency of variables across different equations. Thus, we first focus on a simplified 2-by-2 scenario as a proof-of-concept, with the aim of subsequently generalizing the results to arbitrary scenarios.
\subsection{2-by-2 simplified scenario}\label{ssec:2x2}
Let us consider a simplified scenario where $M=2$ and $K=2$. We can now particularize \eqref{eq:int_ki}, which gives the following system of two equations
\begin{equation}\label{eq:sys_eqs}
    \left\{\begin{aligned} \frac{w_{11}}{w_{21}}&=-\frac{h_{22}(\alpha_2+\beta_2(\vert w_{21} \vert^2+\vert w_{22} \vert^2)}{h_{21}(\alpha_1+\beta_1(\vert w_{11} \vert^2+\vert w_{12} \vert^2)} \\
    \frac{w_{12}}{w_{22}}&=-\frac{h_{12}(\alpha_2+\beta_2(\vert w_{21} \vert^2+\vert w_{22} \vert^2)}{h_{11}(\alpha_1+\beta_1(\vert w_{11} \vert^2+\vert w_{12} \vert^2)}  
     \end{aligned}\right.,
\end{equation}
where $w_{mk}$ is the $(m,k)$th element of $\bs{W}$, and $h_{km}$ is the $(k,m)$th element of $\bs{H}$. Assuming the complex polar form $w_{mk}=\vert w_{mk}\vert\mrm{e}^{\jmath\varphi_{mk}}$, we can rewrite \eqref{eq:sys_eqs} as
\begin{equation}\label{eq:sys_eqs_ph}
    \left\{\begin{aligned} 
    \mrm{e}^{\jmath(\varphi_{11}-\varphi_{21})}&=-\frac{h_{22}(\alpha_2+\beta_2(\vert w_{21} \vert^2+\vert w_{22} \vert^2)\vert w_{21} \vert}{h_{21}(\alpha_1+\beta_1(\vert w_{11} \vert^2+\vert w_{12} \vert^2)\vert w_{11} \vert} \\
    \mrm{e}^{\jmath(\varphi_{12}-\varphi_{22})}&=-\frac{h_{12}(\alpha_2+\beta_2(\vert w_{21} \vert^2+\vert w_{22} \vert^2)\vert w_{22} \vert}{h_{11}(\alpha_1+\beta_1(\vert w_{11} \vert^2+\vert w_{12} \vert^2)\vert w_{12} \vert}    
     \end{aligned}\right.,
\end{equation}
where the \ac{RHS} of the equations only depends on the amplitudes of the unknowns from $\bs{W}$, and the \ac{LHS} depends on two decoupled phase differences associated to these unknowns. We may now observe that \eqref{eq:sys_eqs_ph} always has a explicit solution as long as the \ac{RHS} is uni-modulus since the \ac{LHS} allows fixing the resulting phase for each equation independently. Thus, we should solve instead
\begin{equation}\label{eq:sys_eqs_unimod}
    \left\{\begin{aligned} 
    r_1\big(\{\vert w_{mk}\vert\} \big) \! \triangleq \! \frac{\vert h_{22}(\alpha_2+\beta_2(\vert w_{21} \vert^2+\vert w_{22} \vert^2)\vert \vert w_{21} \vert}{\vert h_{21}(\alpha_1+\beta_1(\vert w_{11} \vert^2+\vert w_{12} \vert^2)\vert \vert w_{11} \vert} \!  =\!1 \\
    r_2\big(\{\vert w_{mk}\vert\} \big)\!\triangleq \!\frac{\vert h_{12}(\alpha_2+\beta_2(\vert w_{21} \vert^2+\vert w_{22} \vert^2)\vert \vert w_{22} \vert}{\vert h_{11}(\alpha_1+\beta_1(\vert w_{11} \vert^2+\vert w_{12} \vert^2)\vert \vert w_{12} \vert} \!=\!1
     \end{aligned}\right..
\end{equation}

\begin{obs}\label{obs:sol_scaling}
    The dominant scaling for the numerator of $r_1\big(\{\vert w_{mk}\vert\})$ is $\mathcal{O}(\vert w_{21}\vert^3)$, while for the denominator it is $\mathcal{O}(\vert w_{11}\vert^3) $, both associated to amplitudes of the entries from the second column of $\bs{W}$. On the other hand, the dominant scaling for the numerator of $r_2\big(\{\vert w_{mk}\vert\})$ is $\mathcal{O}(\vert w_{22}\vert^3)$, while for the denominator it is $\mathcal{O}(\vert w_{12}\vert^3)$, both associated to amplitudes of the entries from the first column of $\bs{W}$.
\end{obs}

Taking into account Observation \ref{obs:sol_scaling}, we may use iterative algorithms to find the amplitude values allowing for a solution to \eqref{eq:sys_eqs_unimod}, while ensuring the power constraint \eqref{eq:pw_cons}. For example, if the amplitude of $r_i(\{\vert w_{mk} \vert\})$ for one of the equations in \eqref{eq:sys_eqs_unimod} is above (below) 1, we may decrease (increase) the power of the element of $\bs{W}$ incurring cubic scaling on the amplitude of the numerator, and increase (decrease) by the same amount the power of the element of $\bs{W}$ incurring cubic scaling on the amplitude of the denominator. This may be performed iteratively until each $r_i(\{\vert w_{mk} \vert\})$ converges to 1 at both equations. Note that the elements of $\bs{W}$ modified for each equation are associated to the same column of $\bs{W}$, allowing us to maintain the power constraint \eqref{eq:pw_cons} throughout the variable updates. These steps are algorithmically described in Algorithm~\ref{alg:nla-zf}.

 \begin{algorithm}
 \caption{\ac{NLA-ZF} algorithm for 2-by-2 case.}
 \label{alg:nla-zf}
 \begin{algorithmic}[1]
 \renewcommand{\algorithmicrequire}{\textbf{Input:}}
 \renewcommand{\algorithmicensure}{\textbf{Output:}}
  \REQUIRE $\bs{H}$, $\{\alpha_m\}$, $\{\beta_m\}$, $\mrm{tol}$, $\epsilon$.
  \ENSURE $\bs{W}$
  \STATE Select initial $\bs{W}$ fulfilling \eqref{eq:pw_cons}.
 \WHILE{$r_1(\{\vert w_{mk}\}), r_2(\{\vert w_{mk}\}) \notin [1-\mrm{tol},1+\mrm{tol}]$}
\IF{$r_1(\{\vert w_{mk}\})<1+\mrm{tol}$}
\STATE $\vert w_{11}\vert = \sqrt{\vert w_{11} \vert^2-\epsilon}$, $\;\;\vert w_{21}\vert = \sqrt{\vert w_{21} \vert^2+\epsilon}$
\ELSIF{$r_1(\{\vert w_{mk}\})>1+\mrm{tol}$}
\STATE $\vert w_{11}\vert = \sqrt{\vert w_{11} \vert^2+\epsilon}$, $\;\;\vert w_{21}\vert = \sqrt{\vert w_{21} \vert^2-\epsilon}$
\ENDIF
\IF{$r_2(\{\vert w_{mk}\})<1+\mrm{tol}$}
\STATE $\vert w_{12}\vert = \sqrt{\vert w_{12} \vert^2-\epsilon}$,
 $\;\;\vert w_{22}\vert = \sqrt{\vert w_{22} \vert^2+\epsilon}$
\ELSIF{$r_2(\{\vert w_{mk}\})>1+\mrm{tol}$}
\STATE $\vert w_{12}\vert = \sqrt{\vert w_{12} \vert^2+\epsilon}$, $\;\;\vert w_{22}\vert = \sqrt{\vert w_{22} \vert^2-\epsilon}$
\ENDIF
\ENDWHILE
\end{algorithmic}
\end{algorithm}

An important limitation of Algorithm~\ref{alg:nla-zf} is that its convergence rate depends largely on $\epsilon$. On the other hand, if $\epsilon$ is too large with respect to the allowed tolerance, it may not be possible to converge at all. Instead, we may consider an alternative fixed-point iteration algorithm with faster convergence by iteratively solving \eqref{eq:sys_eqs_unimod}, assuming fixed $g_i(\{\vert w_{mk} \vert^2\})=r_i(\{\vert w_{mk} \vert\})\frac{\vert w_{1k} \vert}{\vert w_{2k} \vert}$ at each iteration, while enforcing the power constraint \eqref{eq:pw_cons} for each column of $\bs{W}$. Specifically, fixing $g_i(\{\vert w_{mk} \vert^2\})$ with the current entries $\{\vert w_{mk} \vert\}$, we may solve iteratively for each $k\in\{1,2\}$ the system of equations
\begin{equation}
    \left\{\begin{aligned} 
    &\vert w_{1k} \vert^2+ \vert w_{2k} \vert^2 = E_{s,k}\\
    &g_i^2(\{\vert w_{mk} \vert\})\vert w_{2k}\vert^2 = \vert w_{1k} \vert^2    
     \end{aligned}\right. .
\end{equation}
This approach is described in Algorithm~\ref{alg:nla-zf_imp}. Formal convergence criteria may be analyzed in future work, but we have found the algorithm to attain fast convergence in the cases analyzed in Section~\ref{sec:num_res}---taking less than $10$ iterations to reach a tolerance of $10^{-4}$. Note that convergence is guaranteed in the case of weak coupling, i.e., when $\beta_m$ is significantly smaller than $\alpha_m$, which is a reasonable assumption when considering realistic amplifier responses \cite{3GPP_amp}. 

\begin{algorithm}
 \caption{\ac{NLA-ZF} improved algorithm for 2-by-2 case.}
 \label{alg:nla-zf_imp}
 \begin{algorithmic}[1]
 \renewcommand{\algorithmicrequire}{\textbf{Input:}}
 \renewcommand{\algorithmicensure}{\textbf{Output:}}
  \REQUIRE $\bs{H}$, $\{\alpha_m\}$, $\{\beta_m\}$, $\mrm{tol}$.
  \ENSURE $\bs{W}$
  \STATE Start with $n=0$ and select initial $\bs{W}(0)$ fulfilling \eqref{eq:pw_cons}.
 \WHILE{$r_1(\{\vert w_{mk}\}), r_2(\{\vert w_{mk}\}) \notin [1-\mrm{tol},1+\mrm{tol}]$}\vspace{0.3em}
\STATE $\vert w_{21}^{(n+1)}\vert^2 = \frac{E_{s,1}}{1+g_1(\{\vert w_{mk} ^{(n)} \vert\})}$\vspace{0.3em}
\STATE $\vert w_{11}^{(n+1)}\vert^2 = E_{s,1}-\vert w_{21}^{(n+1)}\vert^2$ \vspace{0.3em}
\STATE $\vert w_{12}^{(n+1)}\vert^2 =\frac{E_{s,2}}{1+g_2(\{\vert w_{m1}^{(n+1)}\vert^2,\vert w_{m2}^{(n)} \vert^2\})}$\vspace{0.3em}
\STATE $\vert w_{22}^{(n+1)}\vert^2 = E_{s,2}-\vert w_{12}^{(n+1)}\vert^2$
\STATE n = n+1
\ENDWHILE
\end{algorithmic}
\end{algorithm}

\subsection{Extension to arbitrary (even) number of BS antennas}\label{ssec:ext_M}
Let us consider an extension of the previous scenario where $K=2$ \acp{UE} are now served by a \ac{BS} with arbitrary number of antennas. Given the structure of the problem, and the available solution for the 2-by-2 case, we consider for increased tractability that the number of \ac{BS} antennas is even, i.e., $M=2L$. However, this restriction has essentially no impact for large $M$, while most practical multi-antenna transceivers employ an even number of antennas. We may then rewrite the \ac{LHS} of \eqref{eq:cond_int_canc} as
\begin{equation}\label{eq:M_2L_int}
    \bs{H}\bs{G}(\bs{W})\bs{W} = \sum_{\ell=1}^{L} \bs{H}_{\ell}\bs{G}_{\ell}(\bs{W}_{\ell}) \bs{W}_{\ell},
\end{equation}
where $\bs{H}_{\ell}$ are the $2\times2$ column blocks of $\bs{H}$, $\bs{W}_{\ell}$ are the $2\times 2$ row blocks of $\bs{W}$, and $\bs{G}_{\ell}(\bs{W}_{\ell})$ are the $2\times 2$ diagonal blocks of $\bs{G}(\bs{W})$. We can then realize that each of the sum elements in \eqref{eq:M_2L_int} may be seen as an independent 2-by-2 simplified scenario. Thus, we may use the methods from Section~\ref{ssec:2x2}, namely Algorithm~\ref{alg:nla-zf} and/or \ref{alg:nla-zf_imp}, to force perfect interference cancellation also in this case. Note that this approach disregards some of the available degrees of freedom, since we could potentially cancel off-diagonal elements among sum elements. However, we still get a valid initial solution to our problem, and further generalization may be considered in future work.

In a 2-\ac{UE} scenario, adding more antennas at the \ac{BS} is mainly justified by the possibility to attain greater beamforming gain. However, if we simply apply the methods from the 2-by-2 case in Section~\ref{ssec:2x2} to the summands in \eqref{eq:M_2L_int}, we end with a combination of $L$ diagonal matrices where the diagonal elements may have arbitrary phases. Recall that the methods described in Section~\ref{ssec:2x2} have no restrictions on the resulting $\gamma_k$ values. Hence, in order to have a useful gain from the extra \ac{BS} antennas we need to enforce that the resulting diagonal summands in \eqref{eq:M_2L_int} are combined in phase. The following proposition shows how the phases of the resulting $\gamma_k$ values when solving \eqref{eq:cond_int_canc} can be freely adjusted.

\begin{prop}\label{prop:sol_phase}
Given a $M\times K$ matrix $\bs{W}$ solving \eqref{eq:cond_int_canc}, we have that $\bs{W}_\mrm{new}=\bs{W}\cdot \diag(\mrm{e}^{\jmath\phi_1},\dots,\mrm{e}^{\jmath\phi_K})$ is also a solution to \eqref{eq:cond_int_canc}, $\forall (\phi_1,\phi_2)\in\mathbb{R}^2$. In other words, we can freely shift the phases of $\gamma_k$, $\forall k$, without affecting solvability of \eqref{eq:cond_int_canc}.
\begin{proof}
    As seen in \eqref{eq:BussG}, the dependency of $\bs{G}(\bs{W})$ on $\bs{W}$ comes through squared norms of the rows of $\bs{W}$. An arbitrary phase shift applied to the columns of $\bs{W}$ has no impact on the squared norms of its rows. Hence, $\bs{G}(\bs{W})$ is invariant to arbitrary phase shifts in the columns of $\bs{W}$, so that the phase shifts will only affect the resulting diagonal matrix.
\end{proof}
\end{prop}

Using Proposition~\ref{prop:sol_phase} we may force in-phase combination of the summands of \eqref{eq:M_2L_int}, after applying on each $2\times2$ block the methods from Section~\ref{ssec:2x2}, by simply finding the phase of the resulting $\gamma_{k,\ell}$ values and applying an inverse phase-shift to the respective column $k$ of $\bs{W}_{\ell}$. This way, the $L$ resulting diagonal matrices will have real and positive diagonal entries, ensuring that they are all combined in phase. 

We may note that the power restriction from \eqref{eq:pw_cons} can be arbitrarily distributed throughout the $W_{\ell}$ blocks in \eqref{eq:M_2L_int}, i.e., we may rewrite such restriction as
\begin{equation}
    \sum_{\ell=1}^{L}\Vert\bs{w}_{\ell,k}^\mrm{col}\Vert^2=E_{s,k}, \;\; \forall k \in \{1,2\},
\end{equation}
where $\bs{w}_{\ell,k}^\mrm{col}$ is now the $k$th column of the $2\times2$ block $\bs{W}_{\ell}$. In this work, we assume equal power distribution among the antenna pairs, which is reasonable under the assumption that all antennas receive approximately the same power. Future work may consider this extra degree of freedom for optimizing performance, namely beamforming gain, further.

\section{Numerical Results}\label{sec:num_res}
We next evaluate the average \ac{SINDR} per user of the proposed approach. The \ac{SINDR} for \ac{UE} $k$ may be derived from \eqref{eq:MIMO_Buss} as
\begin{equation}\label{eq:sindr}
\mrm{SINDR}_{k}= \frac{\left\vert\bs{h}_k^\mrm{T}\bs{G}(\bs{W})\bs{w}_{k}^\mrm{col} \right\vert^2}{\sum_{i\neq k}\left\vert\bs{h}_k^\mrm{T}\bs{G}(\bs{W})\bs{w}_i^\mrm{col}\right\vert^2+\bs{h}_k^\mrm{T}\bs{R}_{\bs{\eta \eta}}\bs{h}^\mrm{*}_k+N_0},
\end{equation}
where $\bs{h}_k^\mrm{T}$ corresponds to the $k$th row of $\bs{H}$. Note that the \ac{SINDR} has direct correspondence with the achievable rate through the usual $\log(1+\mrm{SINDR}_k)$ equation since, given $\mathbb{E}\{\bs{\eta}\bs{s}^\mrm{H}\}=\bs{0}$ ensured by the Bussgang theorem, assuming Gaussian $\bs{\eta}$ corresponds to a worse case scenario \cite{worst_noise}.

\begin{figure*}[h]
    \centering
    \begin{subfigure}[b]{0.49\textwidth}
         \centering         \includegraphics[scale=0.55]{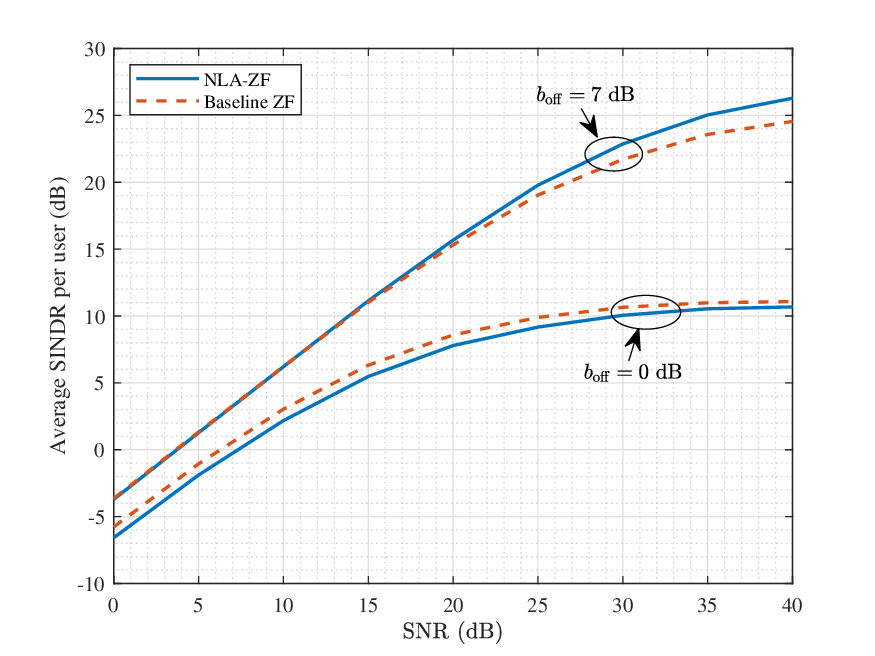}
         \caption{$M=2$}
         \label{fig:M_2}
     \end{subfigure}
     \begin{subfigure}[b]{0.49\textwidth}
         \centering
    \includegraphics[scale=0.55]{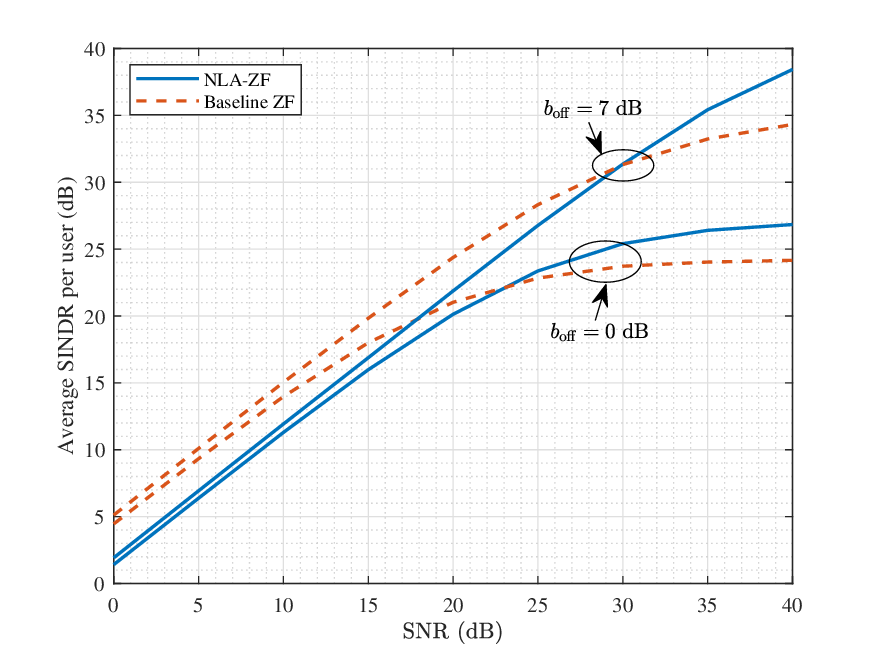}
    \caption{$M=8$}
    \label{fig:M_8}
     \end{subfigure}
      \vspace{-0.5em}
    \caption{Average \ac{SINDR} per user for $K=2$ scenario.}
    \label{fig:rates}
    \vspace{-0.5em}
\end{figure*}

In Fig.~\ref{fig:rates} we plot the average \ac{SINDR} per user of the proposed \ac{NLA-ZF} for the $K=2$ \acp{UE} scenario throughout $10^4$ realizations of an IID fading channel with equal normalized power per entry. The \acp{PA} have been modeled according to the measurements from \cite{3GPP_amp} for the GaN amplifier operated at 2.1 GHz, but they have been fitted to the third-order model as in \cite{lis_amp_dist}. We have included a $\pm10 \%$ uniform random deviation to the non-linearity parameters, independent throughout antennas, to account for the tolerances of the hardware components. The \ac{NLA-ZF} precoder has been characterized using Algorithm~\ref{alg:nla-zf_imp}, with $\mrm{tol}=10^{-4}$, due to faster convergence in the considered cases than Algorithm~\ref{alg:nla-zf}, which may still attain essentially the same performance for small enough $\epsilon$. As a baseline, we have included a naive \ac{ZF} precoder, which is directly obtained by performing the right pseudoinverse of the channel matrix $\bs{H}$, followed by column normalization to account for \eqref{eq:pw_cons}. Note that the baseline \ac{ZF} is assumed to be completely unaware of the \ac{PA} responses. We have also considered results with amplifier back-off, $b_\mrm{off}$, to see the effect of operating the amplifier in a more linear region, hence reducing the influence of the distortion term.

The results for the $M=2$ case in Fig.~\ref{fig:M_2} show that the proposed \ac{NLA-ZF} allow for performance improvements in the regimes where residual interference is significant with respect to distortion and noise, i.e., for higher \ac{SNR} regimes under reasonable back-off so that the distortion power does not dominate the \ac{SINDR} denominator in \eqref{eq:sindr}. For the case $M=8$ in Fig.~\ref{fig:M_8}, we see that the extended method from Section~\ref{ssec:ext_M} attains a reasonable beamforming gain compared to the $M=2$ scenario, but there is a slight degradation compared to the baseline \ac{ZF}. This gap may be reduced by considering more elaborate schemes to distribute the power among the different $\bs{W}_{\ell}$ blocks in \eqref{eq:M_2L_int}. Nevertheless, \ac{NLA-ZF} still outperforms baseline \ac{ZF} as the \ac{SNR} increases and the residual interference gains significance. Moreover, the proposed \ac{NLA-ZF} seems to attain a lower distortion floor than naive \ac{ZF} as we increase the \ac{BS} antennas, since Fig.~\ref{fig:M_8} shows some performance improvement in the distortion dominant regime for $0$ dB back-off. We have also confirmed these findings by evaluating \ac{SIR} and \ac{SDR} values in the scenarios from Fig.~\ref{fig:rates}, which are fairly constant within the considered \ac{SNR} range. These are reported in Table~\ref{table:SIR_SDR}, where we have averaged the small fluctuations with respect to \ac{SNR}. Note that the \ac{SIR} for the considered \ac{NLA-ZF} approach may be further increased by considering a lower tolerance in the algorithm.

\begin{table}[h]
\caption{\ac{SDR} and \ac{SIR} values.}
\begin{tabular}{ |c|c|c|c|c|c| }
\cline{3-6}
\multicolumn{2}{c|}{} &  \multicolumn{2}{|c|}{$M=2$} & \multicolumn{2}{|c|}{$M=8$} \\
 \cline{3-6}
 \multicolumn{2}{c|}{} & \textbf{SIR} & \textbf{SDR} & \textbf{SIR} & \textbf{SDR} \\ 
\cline{3-6}
 \hline
 \multirow{2}{*}{\textbf{\ac{NLA-ZF}}} & $b_{\mrm{off}}=0$ dB & 139 dB & 11 dB & 133 dB & 27 dB\\
  \cline{2-6}
 & $b_{\mrm{off}}=7$ dB & 163 dB & 28 dB & 156 dB & 42 dB \\
 \hline
 \multirow{2}{*}{\textbf{\ac{ZF}}} & $b_{\mrm{off}}=0$ dB & 40 dB & 12 dB & 43 dB & 25 dB \\
  \cline{2-6}
 & $b_{\mrm{off}}=7$ dB & 49 dB & 28 dB & 45 dB & 40 dB \\
 \hline
\end{tabular}
\label{table:SIR_SDR}
\end{table}

\section{Conclusions}
We have shown that \ac{PA} non-linearties can limit the interference cancellation properties of \ac{ZF} linear precoding in downlink \ac{MU-MIMO}. In such scenarios, we have studied how to design a linear precoding scheme, namely \ac{NLA-ZF}, that employs knowledge of the \ac{PA} amplifier response to attain perfect interference cancellation. We have derived two algorithms that achieve \ac{NLA-ZF} in the two-\ac{UE} scenario under even number of \ac{BS} antennas. The proposed approach may attain \ac{SINDR} gains over traditional \ac{ZF}, especially when residual interference becomes significant. Future work may consider generalizing the results to achieve \ac{NLA-ZF} precoding with more than two \acp{UE}, which is still an open problem.
\bibliographystyle{IEEEtran}

\bibliography{IEEEabrv,wax}
\balance

%
\IEEEpeerreviewmaketitle

\end{document}